\documentclass[conference]{IEEEtran}
\IEEEoverridecommandlockouts
\usepackage{cite}
\usepackage{amsmath,amssymb,amsfonts}
\usepackage{algorithmic}
\usepackage{graphicx}
\usepackage{textcomp}
\usepackage{xcolor}
\def\BibTeX{{\rm B\kern-.05em{\sc i\kern-.025em b}\kern-.08em
    T\kern-.1667em\lower.7ex\hbox{E}\kern-.125emX}}

\usepackage{makecell}
\usepackage{amsmath}
\usepackage{amsmath, amssymb, amscd, amsthm, amsfonts}
\usepackage{colortbl}
\usepackage{soul}

\usepackage[colorlinks=true,linkcolor=blue,anchorcolor=black,citecolor=red,filecolor=black,menucolor=black,runcolor=black,urlcolor=black]{hyperref}
    
\begin{document}

\title{Machine Learning Based Cyber System Restoration for IEC 61850 Based Digital Substations\\

\author{\IEEEauthorblockN{Kuchan Park, Mansi Girdhar, Junho Hong, Wencong Su}
\IEEEauthorblockA{\textit{\small Department of Electrical and Computer Engineering} \\
\textit{\small University of Michigan-Dearborn}\\
\small Dearborn, MI, USA \\
\small kuchan@umich.edu, gmansi@umich.edu, jhwr@umich.edu, wencong@umich.edu} 
\and
\and
\IEEEauthorblockN{
Akila Herath, Chen-Ching Liu}
\IEEEauthorblockA{\textit{\small Department of Electrical and Computer Engineering} \\
\textit{\small Virginia Tech}\\
\small Blacksburg, VA, USA \\
\small akilaasansana@vt.edu, ccliu@vt.edu}
}


}



\maketitle

\vspace{-10em}  

\begin{abstract}
Substation Automation Systems (SAS) that adhere to the International Electrotechnical Commission (IEC) 61850 standard have already been widely implemented across various on-site local substations. However, the digitalization of substations, which involves the use of cyber system, inherently increases their vulnerability to cyberattacks. This paper proposes the detection of cyberattacks through an anomaly-based approach utilizing Machine Learning (ML) methods within central control systems of the power system network. Furthermore, when an anomaly is identified, mitigation and restoration strategies employing concurrent Intelligent Electronic Devices (CIEDs) are utilized to ensure robust substation automation system operations. The proposed ML model is trained using Sampled Value (SV) and Generic Object Oriented Substation Event (GOOSE) data from each substation within the entire transmission system. As a result, the trained ML models can classify cyberattacks and normal faults, while the use of CIEDs contributes to cyberattack mitigation, and substation restoration.
\end{abstract}


\section{Introduction}
The transmission system is essential infrastructure for delivering power with minimal losses. Protecting this transmission network is critical for maintaining the stability, reliability, and resilience of the entire power system, including distribution and generation systems~\cite{mazibuko2024review}. Protection schemes, such as overcurrent, differential, and distance protection using Intelligent Electronic Devices (IEDs), have been widely adopted across numerous countries and regions. These schemes contribute to stable transmission systems with high accuracy and rapid response times. Those are categorized as rule-based protection schemes, typically executed according to predefined threshold values. However, they have the drawback of being less effective in detecting and responding to cyberattacks targeting digitalized power systems. Therefore, data-driven methods for distinguishing power system events are advantageous for detecting anomalies, leading to active research into anomaly-based approaches that rely on data itself~\cite{bhattacharya2024ml}. In addition to detecting cyberattacks, developing strategies to respond to them is a crucial aspect of this topic. A lack of adequate response to malicious attacks can lead to prolonged blackouts, critical equipment damage, and significant economic repercussions. Therefore, strategies for mitigation and restoration against anomalies in substation operations are also essential functions.

Recent studies have been conducted on this topic. The study by~\cite{bhattacharya2024ml} defines cyberattacks according to the MITRE Adversarial Tactics, Techniques, and Common Knowledge (MITRE ATT\&CK) framework and detects anomalies using a machine learning-based approach. ML models are trained utilizing the SV and GOOSE information from the IEC 61850 communication protocol. Methods for securing Routable SV (R-SV) and Routable GOOSE (R-GOOSE) were proposed by~\cite{ustun2020implementing}. The authors of~\cite{park2024machine} trained ML models using Root Mean Square (RMS) measurements from each substation within the transmission system. The research by~\cite{hong2022automated} developed the cybersecurity scheme for the local substation. The work by~\cite{habib2017adaptive} structured mitigation strategies for protection systems by incorporating Battery Energy Storage System (BESS) and Hybrid Energy Storage (HES), which are components within the microgrid. A substation system restoration scheme algorithm based on resilience characteristics was proposed by~\cite{rahiminejad2023resilience}. These studies either focus solely on evaluating the performance of data-driven based ML models for detecting anomalies or emphasize only on mitigation and restoration strategies when abnormalities occur. However, detecting cyberattacks and subsequently implementing mitigation and recovery strategies are continuous events that must be considered together to safeguard substation automation systems. In addition, the anomaly-based detection methods used in these studies face challenges in accurately identifying specific locations and fault types. This can provide field technicians with precise information to expedite the repair of transmission line failures, serving as a valuable guideline. As a result, implementing these methods across the wider scope of the entire power system presents significant challenges, underscoring the necessity for solutions that are both more extensive and adaptable.

Accordingly, in order to overcome the challenges, this study presents an approach to fault and cyberattack detection using machine learning algorithms, combined with mitigation and restoration strategy. The proposed ML based anomaly detection framework enables operators to perform post event studies more efficiently compared to traditional approaches by considering cyber system data and physical system data. Additionally, the performance of the mitigation and restoration strategy was initially validated by using the CIED. The methodology was tested and validated using the real-time power system simulator. The test results showed that the proposed approach effectively distinguished between cyberattacks and conventional power system faults, including identifying their locations. Furthermore, this testbed demonstrated that transitioning from IEDs to CIEDs effectively mitigates cyberattacks and restores functionality within the substation. Key contributions of this paper are as follows:
\begin{itemize}
    \item The proposed method uses machine learning to analyze physical system data (e.g., voltage, angle, and frequency in SV packets) and digital system data (e.g., circuit breaker (CB) trip signals and CB status data in GOOSE packets) from both the affected substation and adjacent substations. This information is sent to the control center system, such as Supervisory Control and Data Acquisition (SCADA). Within SCADA, trained ML models are used to determine whether the current protection action was triggered by a cyberattack or by a normal fault.
    \item Without a mitigation and restoration strategy, detecting a cyberattack alone is insufficient to prevent adverse impacts on the power system. However, the proposed method allows the CIED to rapidly take over the functions of the compromised IEDs when the substation is under attack, thereby preventing damage to the power system. 
\end{itemize}
The remaining part of this paper is organized as follows: Section II illustrates a methodology. In Section III, case studies are explained. Finally, this paper is concluded in Section IV.

\section{METHODOLOGY}

\setlength{\textfloatsep}{5pt}
\begin{figure}[!t]
\centerline{\includegraphics[width=1.0\linewidth]{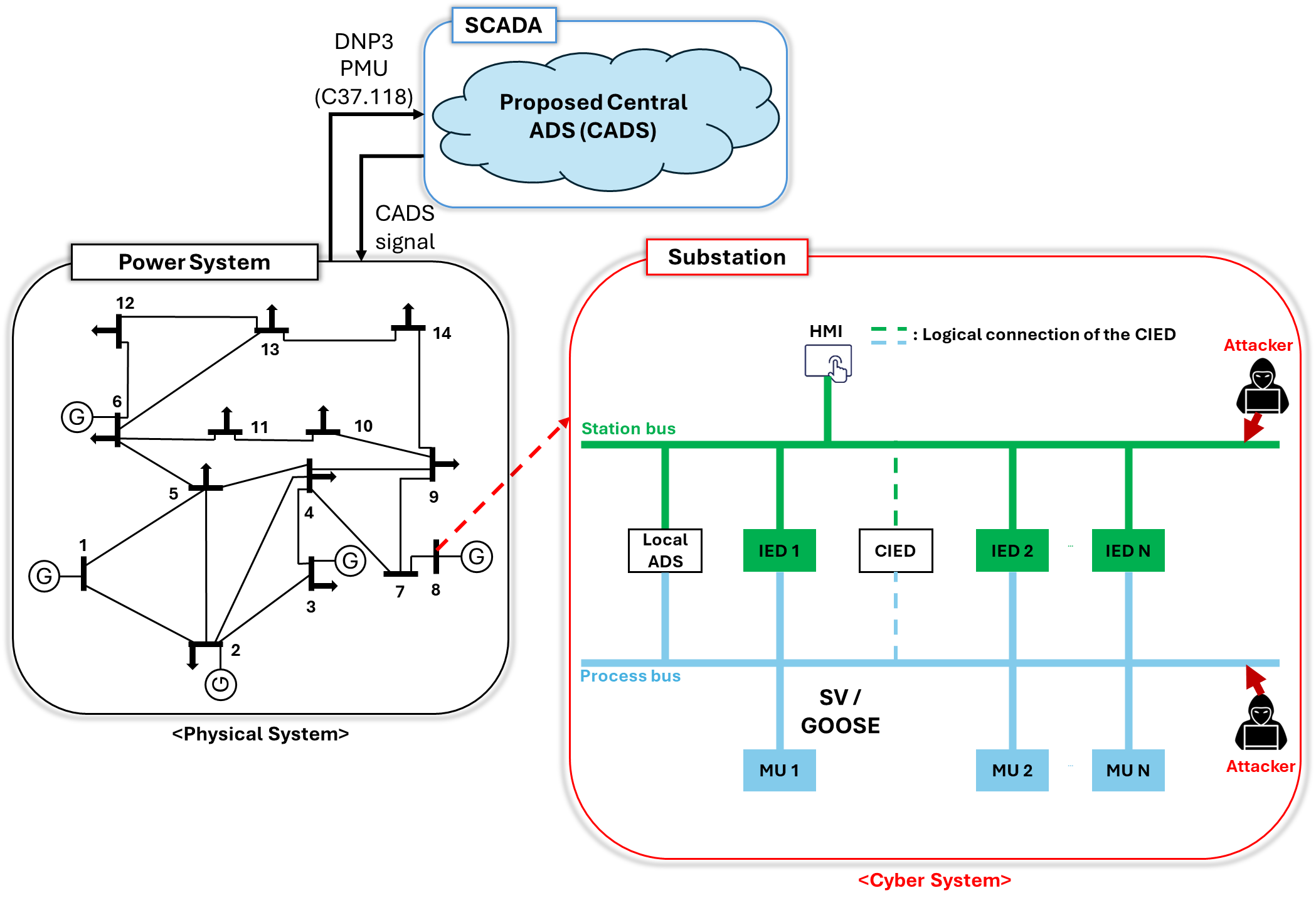}}
\caption{The proposed cyber system restoration framework for digital substations.}
\label{fig1}
\end{figure}

Fig. \ref{fig1} provides an overview of the proposed method in this study. The example of the digital substation at bus 8 is shown in an enlarged form. Same structures exist at each of the 14 buses. Through the process bus, SV packets carry analog measurements of voltage and current from MUs to IEDs. The GOOSE packet includes information regarding the trip signal and CB status. Phasor Measurement Unit (PMU) data is used to send the substation measurement to the control center. In this paper, we propose a method for detecting anomalies, such as cyberattacks, in the control center’s anomaly detection system based on information sent from substations. Typically, the protection actions using the IED are executed within 1--3 cycles, while anomaly detection in the SCADA is completed within 1--3 seconds. In other words, once the IED initiates a protection action, SCADA gathers information from all substations to assess whether the action was triggered by a cyberattack or a normal fault. If the action is identified as cyberattack induced, the CIED fully assumes the original IED functions, isolating the compromised IEDs. As a result, the substation protection system continues to operate normally without any abnormal actions. Through this sequence, the proposed detection, mitigation, and restoration processes are achieved.

\subsection{Proposed Anomaly Detection Method}

\begin{figure}[!t]
\centerline{\includegraphics[width=0.9\columnwidth]{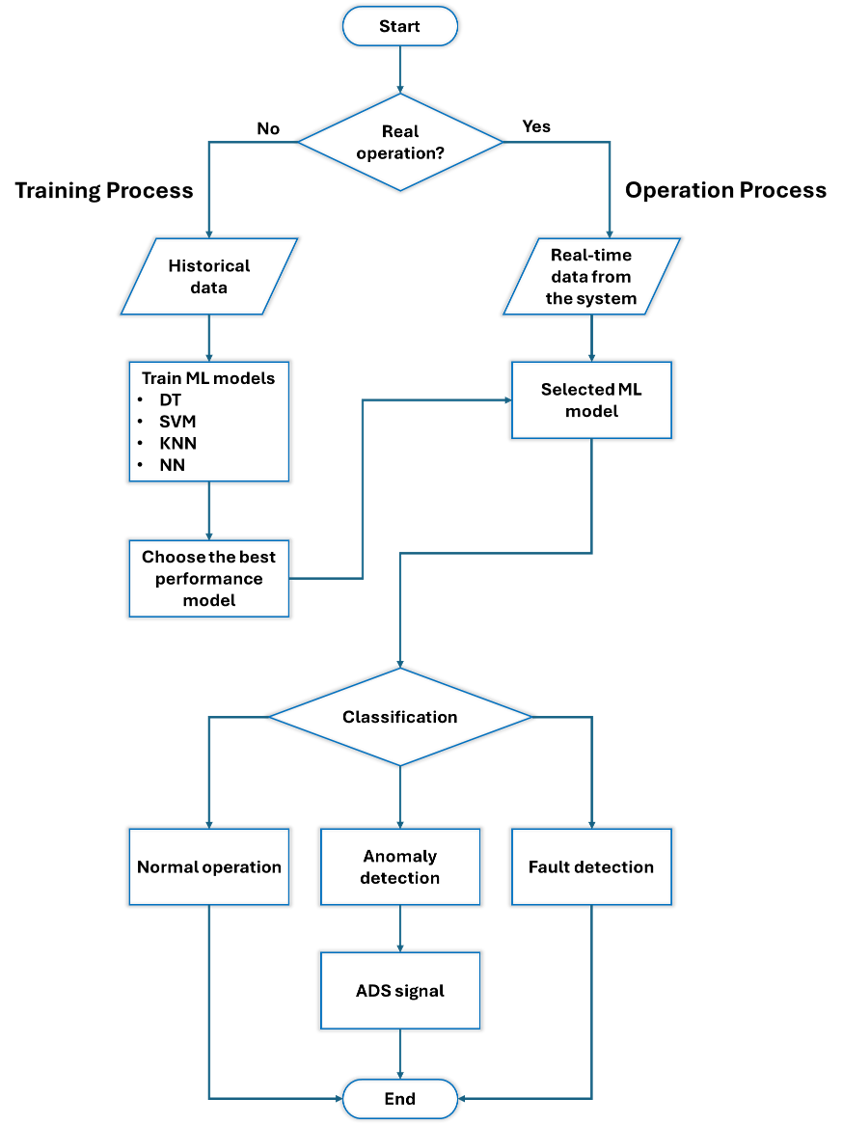}}
\caption{The flowchart of proposed anomaly detection method.}
\label{fig2}
\end{figure}

Fig. \ref{fig2} outlines the flowchart for the proposed anomaly detection approach, structured into two primary stages. The left section of the flowchart represents the process for training various ML models and selecting the most effective model among them. The right section displays the real-time analysis process where the trained ML model is utilized to assess real operational events. Based on the results of the proposed ML-based ADS in the real-time operation, the current situation is classified as either normal operation, a fault condition, or a cyberattack. If a cyberattack is detected, the ADS sends a signal from SCADA to the substation, initiating the proposed mitigation and cyber restoration strategy.

As shown in Table \ref{T1}, this study categorizes operational conditions into 12 distinct classes. The first class (Class 0) represents a normal, fault-free operating condition without any abnormal events. Classes 1 through 10 correspond to ten unique types of physical faults, while Class 11 is specifically designated for cyberattack incidents. Table \ref{T2} provides details on the features used by the machine learning model, including voltage, phase, and frequency values. Additionally, CB trip signals and CB status information are simultaneously considered.

To rigorously test the IED’s capability to discern genuine faults from cyber intrusions, a strategy was employed using authentic fault voltage data for SV faults. Additionally, the cyberattack classification includes scenarios in which an attacker gains access to the substation network to manipulate GOOSE messages. A key aspect of the proposed strategy is its simultaneous consideration of both cyber and physical features. For example, if the voltage value (a physical data point obtainable from SV data) indicates a fault condition, yet the CB trip signal (a cyber data point from GOOSE packets) does not activate, this could signal an anomaly or potential attack. This approach benefits from examining the correlation between cyber and physical data concurrently.

Another significant facet of the proposed strategy is that these cyberattacks are confined to a single substation, meaning no fault transients are observable on adjacent buses, which enhances the clarity in distinguishing them from genuine faults.

\begin{figure}[t]
\centering
\centerline{\includegraphics[width=1.0\columnwidth]{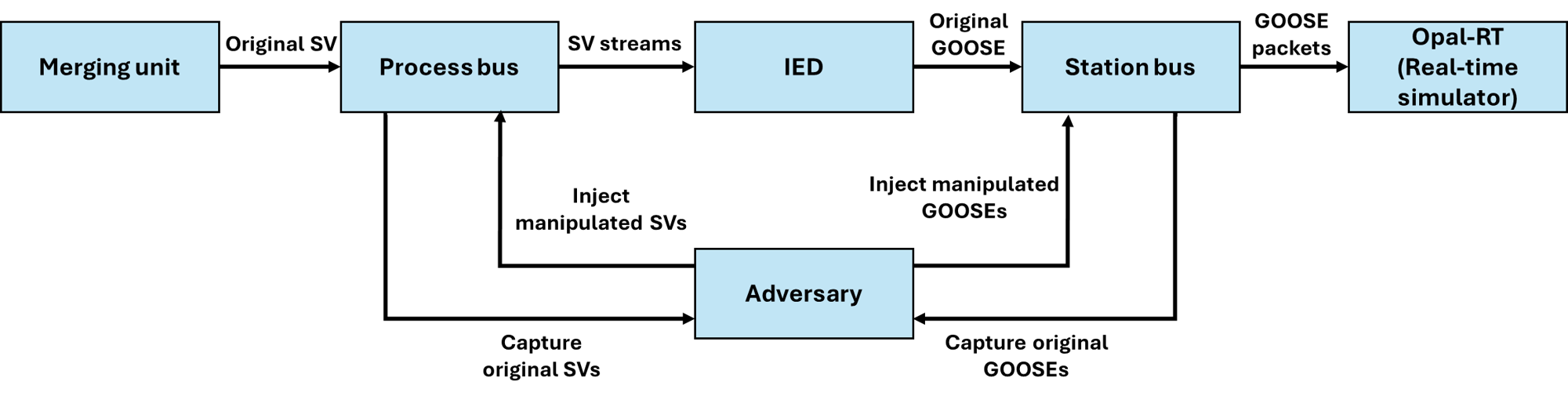}}
\caption{An example of cyberattacks for SV and GOOSE messages.}
\label{fig3}
\end{figure}

In the cyberattack scenario, both replay and False Data Injection (FDI) methods were tested, as illustrated in Fig. \ref{fig3}. Replay attacks, in particular, involve retransmitting previously intercepted SV and GOOSE packets containing fault currents, voltage signals, and CB trip signals. To execute a replay attack, an attacker may access the monitoring port of the process bus or station bus Ethernet switch to capture critical SV and GOOSE message data. If successful, this tactic triggers the protection function of IEDs, leading to the operation of circuit breakers.

\setlength{\textfloatsep}{5pt}
\begin{table}[]
\centering
\caption{Classes for the proposed anomaly detection}
\label{T1}
\begin{tabular}{|c|c|}
\hline
\textbf{Class index} & \textbf{Description} \\ \hline
0                    & Normal operation     \\ \hline
1                    & A-gnd fault          \\ \hline
2                    & B-gnd fault          \\ \hline
3                    & C-gnd fault          \\ \hline
4                    & AB-gnd fault         \\ \hline
5                    & BC-gnd fault         \\ \hline
6                    & AC-gnd fault         \\ \hline
7                    & A-B fault            \\ \hline
8                    & B-C fault            \\ \hline
9                    & A-C fault            \\ \hline
10                   & ABC-gnd fault        \\ \hline
11                   & Cyberattack          \\ \hline
\end{tabular}
\end{table}

\begin{table}[]
\setlength{\tabcolsep}{3pt}
\centering
\caption{Features of the ML models}
\label{T2}
\begin{tabular}{|c|c|l|}
\hline
\textbf{Category} & \textbf{\begin{tabular}[c]{@{}c@{}}Number\\ of features\end{tabular}} & \multicolumn{1}{c|}{\textbf{Description}}                                                                                                                           \\ \hline
Voltage           & 52                                                                    & (3 phases) x (14 buses)                                                                                                                                             \\ \hline
Phase             & 52                                                                    & (3 phases) x (14 buses)                                                                                                                                             \\ \hline
Frequency         & 52                                                                    & (3 phases) x (14 buses)                                                                                                                                             \\ \hline
CB trip signal    & 56                                                                    & \begin{tabular}[c]{@{}l@{}}\{(2 CBs per a line) x (20 lines)\} +\\ \{(1 CB per a generator) x (5 generators)\} +\\ \{(1 CB per a load ) x (11 loads)\}\end{tabular} \\ \hline
CB status         & 56                                                                    & \begin{tabular}[c]{@{}l@{}}\{(2 CBs per a line) x (20 lines)\} +\\ \{(1 CB per a generator) x (5 generators)\} +\\ \{(1 CB per a load ) x (11 loads)\}\end{tabular} \\ \hline
\end{tabular}%
\end{table}

\subsection{Machine Learning Models for Proposed Method}

The proposed structure involves training multiple ML models and selecting the one with the highest accuracy and the lowest false negative. Consequently, a variety of representative ML algorithms were incorporated.

The first model is the Decision Tree (DT), which classifies samples based on their features~\cite{10015734}. Named for its tree structure, a decision tree functions by asking a series of questions about attribute values to reach the correct answer or class. The endpoint or solution to each question is known as a node, and the connections between nodes are referred to as edges. DTs primarily use either the Gini index or entropy as a loss function. In the proposed method, the Gini index is applied as follows.

\begin{equation} \label{E1}
    gini(D)=1-\sum_{j=1}^{n}p_{j}^{2},
 \end{equation}

 where $D$ represents the number of data, $n$ means the number of classes, $p$ donates the ratio of the data belonging to class $j$.

The second model is the Support Vector Machine (SVM), which seeks the optimal boundary, or hyperplane, to separate different classes~\cite{10329926}. In this case, a Gaussian SVM is utilized to effectively classify data with non-linear characteristics.

 \begin{equation} \label{E2}
    \underset{\alpha }{max}\sum_{i=1}^{m}\alpha _{i}-\frac{1}{2}\sum_{i=1}^{m}\; \sum_{j=1}^{m}\alpha _{i}\alpha _{j}y_{i}y_{j}K(x_{i}\cdot x_{j}).
 \end{equation}
 \begin{equation*} \label{E2-1}
   subject\: to\: \alpha _{i}\geq 0,\: \sum_{i=1}^{m}\alpha _{i}y_{i}=0.
 \end{equation*}

Equation 2 represents the objective function for a standard SVM. Here, $\alpha$ represents the Lagrange multiplier, $m$ is the number of data points, $(x, y)$ denotes each data point, and $K$ refers to the Gaussian kernel.

The third model applied is the K-nearest neighbors (KNN), which predicts outputs based on the K nearest neighbors~\cite{peterson2009k}. KNN employs a distance metric to determine which input data is most similar to the trained data. For KNN, the Minkowski distance function is used to calculate the similarity.

 \begin{equation} \label{E3}
   dist(x,y)=\left ( \sum_{i=1}^{n}\left| x_{i}-y_{i}\right|^{p} \right )^{1/p},
 \end{equation}

where $x,y$ indicate data points, $n$ represents the dimension, and $p$ is the order of the norm.

Finally, a NN is trained. The NN is a type of computing system that organizes patterns to mimic the behavior of neurons in the human brain~\cite{abdi1994neural}.

\subsection{Mitigation \& Restoration Strategy}

Fig. \ref{fig1} presents a diagram illustrating the mitigation and restoration strategy proposed in this research to counter cyberattacks. When an anomaly is detected by the proposed ML model-based ADS, the SCADA system changes the boolean signal indicating a cyberattack from 0 to 1. Upon receiving this signal, the substation fully isolates the compromised IEDs to mitigate the cyberattack. For instance, if IED 1 in Fig. \ref{fig1} is compromised, IED 1 is completely isolated, and the CIED assumes all of its functions. This is accomplished through Software-Defined Networking (SDN) switches in the cyber system \cite{10688802}. Specifically, the SDN switches on the process bus and station bus disable the ports connected to IED 1 while simultaneously enabling the ports connected to the CIED. This approach allows the substation to maintain normal operation with minimal disruption. These procedures follow a predefined rule-based method for controlling Ethernet ports via the SDN switch.

\section{CASE STUDIES}

\subsection{The Performance of ML Models}

The hardware-in-the-loop (HIL) testbed was developed to identify the optimal ML model, utilizing a real-time simulator (e.g., Opal-RT), IEDs, SDN switches, and ADS. For data generation, each of the 20 lines was assigned four distinct fault locations and four different fault impedances, simulating a total of ten fault types. Cyberattack scenarios were generated for each condition, resulting in a comprehensive dataset of 6,400 samples. The real-time input data for the ML model included voltage, angle, frequency, CB trip signal, and CB status information from all buses. This structured approach allowed for an in-depth evaluation of each ML model's ability to detect and respond to specific fault conditions within the power system. For optimal ML model selection, the dataset was divided with 90\% allocated for training and the remaining 10\% for validation. This split ensured a robust assessment by balancing an extensive training set with a sufficiently sized validation set to evaluate each model’s generalization and effectiveness.

Table \ref{T3} provides a comparative summary of the performance metrics for four distinct ML models. While DT and SVM models demonstrated similar accuracy rates, SVM exhibited higher precision, recall, and F1-score values. The KNN model, in comparison, showed lower values in these metrics than SVM, except for accuracy, partly due to DT and KNN models’ tendency to incorrectly classify normal operations as cyberattack instances. In contrast, SVM more accurately distinguished between normal conditions and cyberattack scenarios, resulting in improved performance across most metrics. Notably, the NN model outperformed all others, achieving nearly 99\% across all metrics. Given its superior performance, the NN model was ultimately selected for real-time implementation.

\begin{table}[]
\centering
\caption{Model performance of ML models}
\label{T3}
\begin{tabular}{c|c|c|c|c|}
\cline{2-5}
 & \textbf{Accuracy(\%)} & \textbf{Precision(\%)} & \textbf{Recall(\%)} & \textbf{F1-score(\%)} \\ \hline
\multicolumn{1}{|c|}{\textbf{DT}}  & 95.75 & 90.39 & 90.30 & 90.32 \\ \hline
\multicolumn{1}{|c|}{\textbf{SVM}} & 96.8 & 97.32 & 95.35 & 96.32 \\ \hline
\multicolumn{1}{|c|}{\textbf{KNN}} & 98.75 & 91.41 & 91.62 & 91.50 \\ \hline
\multicolumn{1}{|c|}{\textbf{NN}} & 99.75  & 98.85   & 98.84   & 99.10   \\ \hline
\end{tabular}
\end{table}

\subsection{The Validation of Proposed Method}


\setlength{\textfloatsep}{5pt}
\begin{figure}[!t]
\centering
\centerline{\includegraphics[width=1.0\columnwidth]{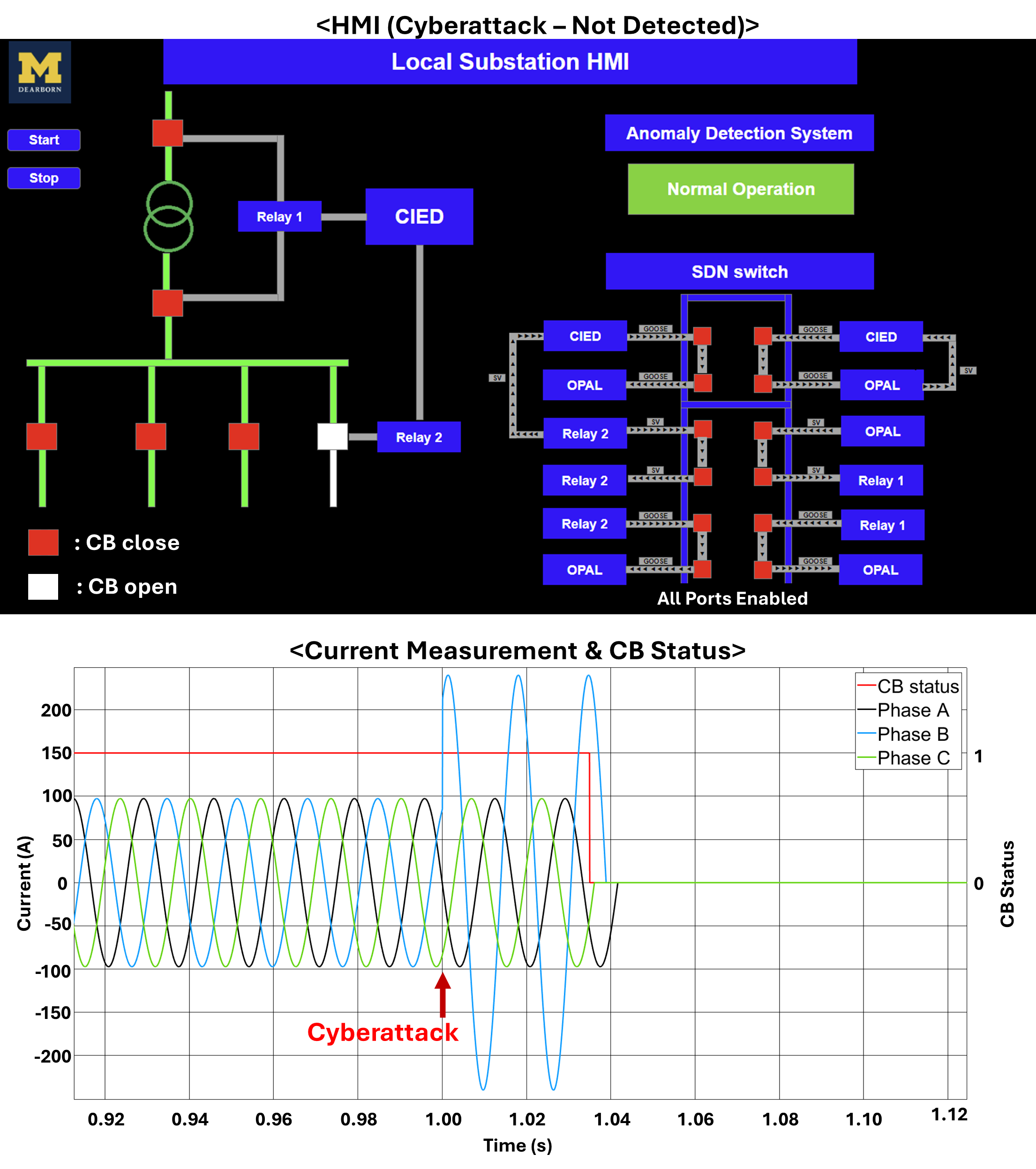}}
\caption{Cyberattack scenario.}
\label{fig5}
\end{figure}

\begin{figure}[!t]
\centering
\centerline{\includegraphics[width=1.0\columnwidth]{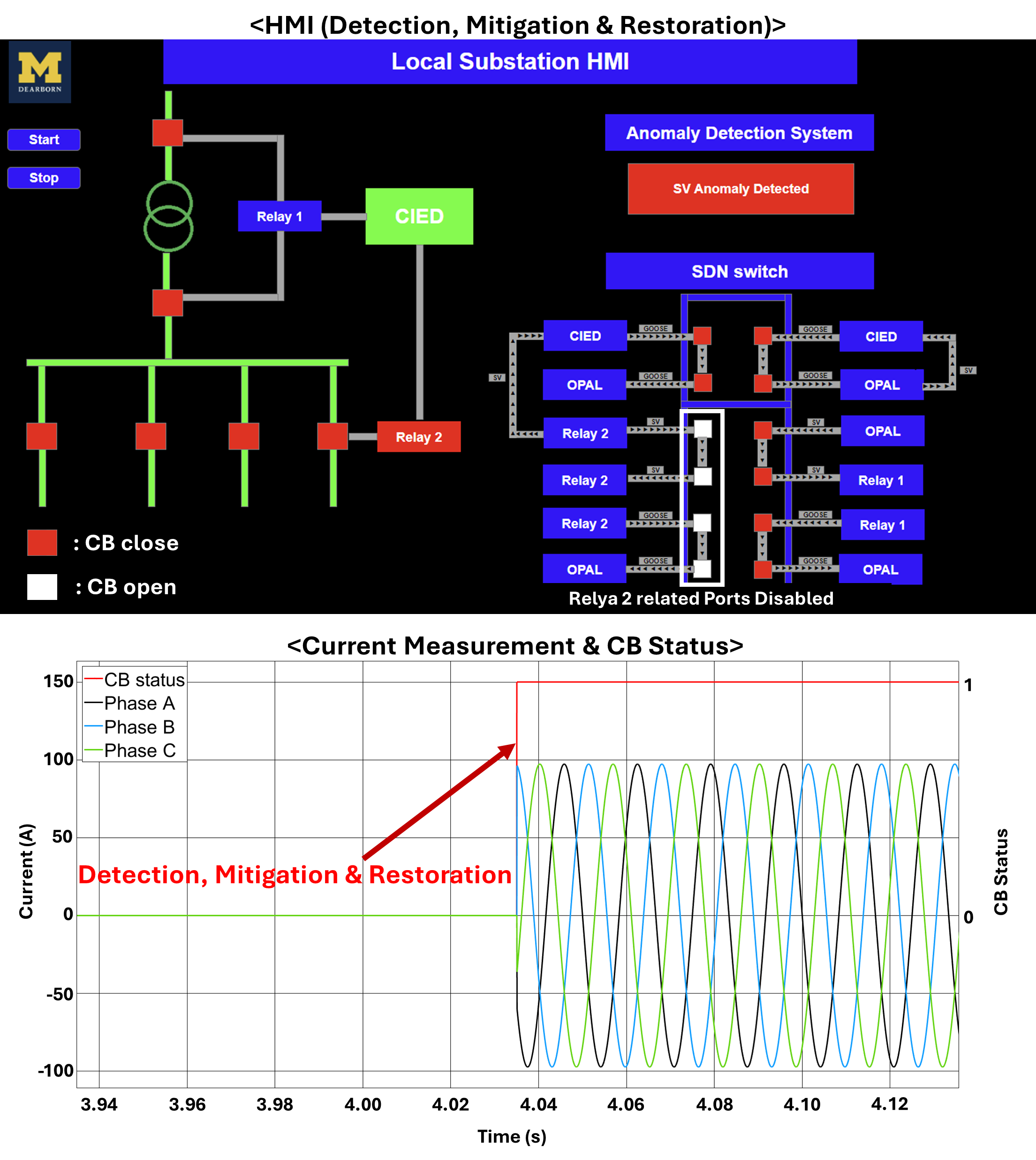}}
\caption{The validation of proposed method.}
\label{fig6}
\end{figure}


The validation results of the proposed method are presented in Fig. \ref{fig5} and Fig. \ref{fig6}. The validation was conducted through the following sequence of scenarios:

\begin{enumerate} \item Normal operation: The substation is assumed to be located at bus 8, with four feeders connecting to the distribution network. The local substation primarily relies on its own independent ADS, unlike the proposed method, which adopts a holistic view of the entire system. Under normal operation, the substation functions without any abnormal behavior.

\item Cyberattack injection: As shown in Fig. \ref{fig5}, a cyberattack is triggered at the 1-second mark. The cyberattack data effectively mimics a B-phase-to-ground fault. At this point, as indicated by the HMI and current measurement graph in Fig. \ref{fig5}, the substation's ADS fails to detect the anomaly, causing the protection system to activate and open the CB within 3 cycles, despite the absence of an actual fault.

\item Application of the proposed method: After the CB opens, SCADA gathers data from all substations across the system over approximately 3 seconds, enabling the proposed ADS to detect the cyberattack. Upon receiving the cyberattack detection signal from SCADA, the substation mitigates the attack by isolating the compromised IEDs and initiates cyber restoration by activating the CIED. In the HMI displayed in Fig. \ref{fig6}, the anomaly is accurately identified, indicating that the CIED has been activated and that the relay 2 connection ports on the SDN switch have been disabled. The graph also shows that normal current values are restored following the cyberattack detection.

\item Performance verification of CIED: To verify the functionality of the CIED, an artificial fault (i.e., a test fault) was introduced after applying the proposed method. The CIED successfully detected the fault, and the CB opened within 3 cycles, confirming that the CIED has fully assumed the functions of relay 2. \end{enumerate}

\section{Conclusion}
 With the increasing decentralization of power systems and advancements in Information \& Communications Technology (ICT), the importance of cybersecurity has become paramount. Responding to this need, this paper introduces an ML-based anomaly detection method aimed at differentiating between cyberattacks and actual faults with precision. Additionally, a strategy for mitigation and restoration is outlined to address detected cyberattacks. The proposed approach was tested and verified using the IEEE 14-bus system.

Looking forward, data from larger transmission networks will be gathered, and a broader range of ML models will be trained to expand the applicability of the proposed method. Further studies will also focus on producing cyberattack input data that closely mirrors real fault conditions to account for worst-case scenarios during ML model training.

\section{Acknowledgment}
\footnotesize
This research was partially funded by the Director of Cybersecurity, Energy Security, and Emergency Response, specifically through the Cybersecurity for Energy Delivery Systems program of the U.S. Department of Energy under contract DE-CR0000021. The views, findings, conclusions, or recommendations presented in this material are solely those of the authors and do not necessarily represent those of the funding sponsors.

\bibliographystyle{IEEEtran}
\bibliography{IEEEabrv,ref}

\end{document}